\documentclass[bm,aps,showpacs,amsfonts,amssymb,twocolumn]{revtex4}  

\usepackage{graphicx}
\usepackage{bm}

\begin{document}
{~}
\title{
Compactified black holes in 
five-dimensional $U(1)^3$ ungauged supergravity
}

\vspace{2cm}

\author{Shinya Tomizawa
}

\vspace{2cm}
\affiliation{
Cosmophysics Group, Theory Center, Institute of Particle and Nuclear Studies,
KEK, Tsukuba, Ibaraki, 305-0801, Japan
}

\begin{abstract} 
We present stationary, nonextremal three charge rotating black hole solutions in the five-dimensional $U(1)^3$ ungauged supergravity. 
At infinity, our solutions behave as a four-dimensional flat spacetime with a compact extra-dimension and hence describe spherical black holes with Kaluza-Klein asymptotics.
\end{abstract}

\date{\today}
\maketitle

Higher dimensional black holes have played an important role in 
understanding basic properties of fundamental theories, such as 
string theory. A number of interesting solutions of higher dimensional, asymptotically flat  
black holes have been discovered recently~\cite{Myers:1986un,Emparan:2001wn,Pom,diring,saturn,Izumi,bi,MishimaIguchi,Gauntlett0,
BMPV,Elvang,EEMR,Elvang3,CY,CLP,CCLP,EEF,Y1,Y2,Y3,Y4}, showing much 
richer structure of their solution space than that of four-dimensional 
black holes. However, since our real world is macroscopically four-dimensional,
extra-dimensions must be compactified in realistic, classical spacetime models. 
The assumption of asymptotic global flatness becomes relevant 
in the context of a certain type of braneworld models or ADD models  
in which size of higher dimensional black holes can be much 
smaller than the curvature radii of the AdS spacetime or the size of compact extra-dimensions.   
Therefore it is of great interest to strudy higher dimensional 
Kaluza-Klein black holes, which look like four-dimensional, at least 
at large distances, while behave as higher dimensions near the event horizon. 
In that context, finding such Kaluza-Klein solutions with compactified extra dimensions and classifying them may also help us to get some insights into the open problem of how to compactify and stabilize extra-dimensions in string theory. In recent years, a lot of Kaluza-Klein black hole solutions have been found and have shown us various physically interesting aspects~\cite{DM,Rasheed,GW,Bena2,Gaiotto,IM,IKMT,NIMT,TIMN,MINT,TI,TYM,GS,BW,tomi}. 
 
\medskip
In this brief report, we construct stationary charged rotating Kaluza-Klein black hole solutions in the bosonic sector of the five-dimensional $U(1)^3$ ungauged supergravity, using so-called {\it squashing transformation}, by which we mean that one first regards the $S^3$ sections of a five-dimensional asymptotically flat black hole spacetime as a fibre bundle of an $S^1$ over an $S^2$, and then performs the deformation that changes the ratio of the radii of the fibre $S^1$ and the base $S^2$. The resultant black hole spacetime behaves as a four dimensional Minkowski spacetime with a compactified 5-the dimension at infinity, while it looks like a five-dimensional spacetime near the horizon. 
Here, applying this deformation to the five-dimensional Cveti\v{c}-Youm~\cite{CY} solutions with equal angular momenta, we obtain the non-extremal Kaluza-Klein black hole solutions with three electric charges and an angular momentum (or a momentum) along the $5$-th dimension in the same theory.

\medskip
The Lagrangian of the bosonic sector of $D=5$ ungauged supergravity is given by
\begin{eqnarray}
&&(-g)^{-1/2}{\cal L}=R-\frac{1}{2}(\partial\varphi_1)^2-\frac{1}{2}(\partial\varphi_2)^2\nonumber\\
&&{\ }-\sum_{i=1}^3\frac{1}{4}X_i^{-2}(F^i)^2+\frac{1}{24}\epsilon_{ijk}\epsilon^{\mu\nu\rho\sigma\lambda}F^i_{\mu\nu}F^j_{\rho\sigma}A^k_\lambda.\label{eq:Lag}
\end{eqnarray}
where $F^i=dA^i$, and
\begin{eqnarray}
&&X_1=e^{-\frac{1}{\sqrt{6}}\varphi_1-\frac{1}{\sqrt{2}}\varphi_2},\quad X_2=e^{-\frac{1}{\sqrt{6}}\varphi_1+\frac{1}{\sqrt{2}}\varphi_2},\\
&&X_3=e^{\frac{2}{\sqrt{6}}\varphi_1}.
\end{eqnarray}

\medskip
Our solutions are written as follows
\begin{eqnarray}
ds^2&=&-\frac{RY}{f_1}dt^2+\frac{r^2R}{Y}k^2dr^2+\frac{1}{4}Rk(\sigma_1^2+\sigma_2^2)\nonumber\\
  &&+\frac{f_1}{4R^2}\left(\sigma_3-\frac{2f_2}{f_1}dt \right)^2,\label{eq:metric}
\end{eqnarray}
\begin{eqnarray}
A^i=\frac{\mu}{r^2H_i}\left[s_ic_idt +\frac{1}{2}l(c_is_js_k-s_ic_jc_k)\sigma_3\right],\label{eq:gauge}
\end{eqnarray}
\begin{eqnarray}
X_i=\frac{R}{r^2 H_i},\label{eq:scalar}
\end{eqnarray}
where left-invariant $1$-forms $\sigma_i$ on $S^3$ are given by
\begin{eqnarray}
&&\sigma_1=\cos\psi d\theta+\sin\psi\sin\theta d\phi,\\
&&\sigma_2=-\sin\psi d\theta+\cos\psi\sin\theta d\phi,\\
&&\sigma_3=d\psi+\cos\theta d\phi,
\end{eqnarray}
and, the functions $(R,H_i,f_1,f_2,Y)$ are
\begin{eqnarray}
R=r^2\left(H_1H_2H_3\right)^{1/3},\quad H_i=1+\frac{\mu s_i^2}{r^2},
\end{eqnarray} 
\begin{eqnarray}
f_1&=&R^3+\mu l^2r^2+\mu^2l^2[2(c_1c_2c_3-s_1s_2s_3)s_1s_2s_3\nonumber\\
    &&-(s_1^2s_2^2+s_2^2s_3^2+s_1^2s_3^2)],\\
f_2&=&\mu l(c_1c_2c_3-s_1s_2s_3)r^2+\mu^2 l s_1s_2s_3,\\
Y&=&r^4-\mu r^2+\mu l^2,
\end{eqnarray}
with the constants $s_i=\sinh\delta_i, c_i=\cosh\delta_i \ (i=1,2,3)$, and the function $k(r)$ is
\begin{eqnarray}
k(r)=\frac{Y(r_\infty)}{(r_\infty^2-r^2)^2}=\frac{r_\infty^4-\mu r_\infty^2+\mu l^2}{(r_\infty^2-r^2)^2},\label{eq:k}
\end{eqnarray}
with the constant $r_\infty$. The ranges of the coordinates $(r,\phi,\theta,\psi)$ are
\begin{eqnarray}
0<r<r_\infty,\ 0\le \phi<2\pi,\ 0\le \theta<\pi,\ 0\le \psi<4\pi.
\end{eqnarray}
In eq.(\ref{eq:gauge}), $(i,j,k)$ take different indices $(i\not= j\not=k\not=i)$. 
Note this solutions can be obtained by replacing the $1$-forms, $(\sigma_1,\sigma_2)$, in the five-dimensional Cveti\v{c}-Youm solution~\cite{CY} of equal angular momenta with the deformed $1$-forms, $(\sqrt{k(r)}\sigma_1,\sqrt{k(r)}\sigma_2)$, and leaving the $1$-form,  $\sigma_3$, unchanged. One can regard each $S^3$ section of a $r$=constant surface as an $S^1$ fibre bundle over an $S^2$ base space. We would like to point out that by this transformation, the ratio of the $S^1$ radii to the $S^2$ radii is drastically changed particularly at infinity, and the round $S^3$ of spatial infinity are squashed. The form (\ref{eq:k}) of the function, $k(r)$, is determined by actually substituting the deformed metric into the Einstein equation and integrating the ordinary derivative equation to the function $k(r)$. One can easily see that the undeformed gauge potentials~(\ref{eq:gauge}) and scalar fields~(\ref{eq:scalar}) are still solutions to the equations 
derived from the action~(\ref{eq:Lag})
by taking the following point into consideration: Under this deformation, $\sqrt{-{\rm det}\ g}$,  $(F^i)^2$ and $(\partial{\varphi_i})^2$  turn out to be transformed into $k(r)^2\sqrt{-{\rm det}\ g}$, $k(r)^{-2}(F^i)^2$, and $k(r)^{-2}(\partial{\varphi_i})^2$,  respectively and therefore, the actions for the Maxwell fields and the scalar fields are both invariant.

\medskip

The horizons are located at the values, $r_\pm=((\mu\pm\sqrt{\mu^2-4\mu l^2})/2)^{1/2}$, of $r$ $(0<r<r_\infty)$ satisfying the equation $Y=r^4-\mu r^2+\mu l^2=0$, and hence 
the necessary and sufficient condition that the spacetime admits two horizons and no closed timelike curve in the domain of outer communication is given by
\begin{eqnarray}
4l^2<\mu<2r_\infty^2,\ 
r_\infty^4-\mu r_\infty^2+\mu l^2>0,\ 
f_1(r_+)>0.
\end{eqnarray}
Note that since the spacetime metric~(\ref{eq:metric}) is analytic outside the outer horizon, the domain of outer communication turns out to be regular everywhere.

\medskip
The squashing transformation cuts off the region $r>r_\infty$ of the spacetime but 
it turns out that $r=r_\infty$ corresponds to spatial infinity since the proper length $\int^r \sqrt{g_{rr}}dr$ diverges in the limit $r \to r_\infty$. 
Therefore,
in order to see the asymptotics of the obtained solutions, we had better define the physically clear, new radial coordinate $\rho$ by
\begin{eqnarray}
\rho=\rho_0\frac{r^2}{r_\infty^2-r^2}
\end{eqnarray}
with the constant
\begin{eqnarray}
\rho_0=\frac{\sqrt{R(r_\infty)Y(r_\infty)}}{2r_\infty^2},
\end{eqnarray}
and, furthermore, introduce the new coordinates $(\bar t,\bar\psi)$ defined by
\begin{eqnarray}
&&dt=\sqrt{\frac{f_1(r_\infty)}{R(r_\infty)Y(r_\infty)}}d\bar t,\\
&&d\psi=d\bar\psi+\frac{2f_2(r_\infty)}{\sqrt{R(r_\infty)Y(r_\infty)f_1(r_\infty)}}d\bar t.
\end{eqnarray}
Then, in terms of these coordinates, the metric at infinity $\rho\to\infty$ $(r\to r_\infty)$ behaves as
\begin{eqnarray}
ds^2&\simeq& -d\bar t^2+d\rho^2+\rho^2(d\theta^2+\sin^2\theta d\phi^2)\nonumber\\
&&+\frac{f_1(r_\infty)}{4R(r_\infty)^2}(d\bar\psi+\cos\theta d\phi)^2+{\cal O}(\rho^{-1}).
\end{eqnarray}
It is clear that the asymptotic metric has the structure of an $S^1$ bundle over the four-dimensional Minkowski spacetime
and the spatial infinity is an $S^1$ fibre bundle over the $S^2$ base space.

\medskip
Next, let us see the relation between the asymptotic charges and the constants appearing in asymptotic form of the metric and gauge potentials. 
The asymptotic charges should be defined as boundary integrals over the spatial infinity. 
Since we are concerned with stationary,
axisymmetric spacetimes with Killing symmetries, the
conserved charges, a mass $M$, an angular momentum $J_\phi$ along the $4$-th dimension, an angular momentum (a momentum) $J_w$ along the $5$-th dimension and electric charges $Q_i$ are computed as follows, 
\begin{eqnarray}
M&=&
-2Lr_\infty^2\rho_0\biggl[\frac{f_1^\prime(r_\infty)}{f_1(r_\infty)}\nonumber\\
&&+\frac{f_2(r_\infty)(R(r_\infty)f_2^\prime(r_\infty)-2f_2(r_\infty)R^\prime(r_\infty))}{R^4(r_\infty)Y(r_\infty)}\nonumber\\
&&-\frac{R^\prime(r_\infty)}{R(r_\infty)}-\frac{Y^\prime(r_\infty)}{Y(r_\infty)}\biggr]\,
\label{def:Mass} ,
\end{eqnarray}
\begin{eqnarray}
J_\phi
=0 \,
\label{def:Ja} ,
\end{eqnarray}
\begin{eqnarray}
J_w
=-\frac{4 r_\infty^2\rho_0(f_2(r_\infty)f_1^\prime(r_\infty)-f_1(r_\infty)f_2^\prime(r_\infty))}{R(r_\infty)^2\sqrt{f_1(r_\infty)R(r_\infty)Y(r_\infty)}} \,
\label{def:Ja},
\end{eqnarray}
\begin{eqnarray}
Q_i
&=&\frac{2\mu \rho_0 L }{H_i(r_\infty)^2r_\infty^2}\sqrt{\frac{f_1(r_\infty)}{R(r_\infty)Y(r_\infty)}}\nonumber\\
&&\times\left(s_ic_i+\frac{2lf_2(r_\infty)(c_is_js_k-s_ic_jc_k)}{f_1(r_\infty)}\right),
\end{eqnarray}
where the prime ${}^\prime$ denotes $\partial_{r^2}$ and $L=2\sqrt{g_{\psi\psi}(r_\infty)}$.

\medskip
Finally, we consider various limits to known solutions in the five-dimensional supergravity.

(i) the limit to the ungauged  minimal supergravity: 
As is well known, when $A_1=A_2=A_3$, the action coincides with that of the bosonic sector in the five-dimensional ungauged minimal supergravity, {\it i.e.}, the Einstein-Maxwell-Chern-Simons theory with a certain value of the coupling constant. For our solutions, this can be achieved by setting $\delta:=\delta_1=\delta_2=\delta_3$, and in fact redefining the parameters by $r^2 \to r^2-\mu s^2$, $\mu\frac{s^2+c^2}{2}\to m$, $\mu sc \to q$ and $l(c-s)\to a$ ($s=\sinh\delta,\ c=\cosh\delta$), one can show the solutions exactly coincide with the nonextremal charged rotating Kaluza-Klein black hole solutions~\cite{NIMT} in the five-dimensional minimal supergravity.

(ii) the limit to asymptotically flat solutions: In the limit of $r_\infty \to \infty$ with the other parameters fixed, we can see $k(r) \to 1$ and hence can derive the five-dimensional asymptotically flat Cveti\v{c}-Youm solutions~\cite{CY} with equal angular momenta. In particular, this solutions admit the limit to the supersymmetric BMPV solutions~\cite{BMPV}.

(iii) the extremal limits:
Taking the limit of $\mu \to 0$, $l\to 0$, $\delta_i \to -\infty$ with keeping $\mu s_ic_i$ and $l(c_i-s_i)$ finite, we have
\begin{eqnarray}
ds^2&=&-(Z_1Z_2Z_3)^{-\frac{2}{3}}\left(dT+\frac{\mu lc_1c_2c_3}{R(r_\infty)}V^{-1}\sigma_3\right)^2\nonumber\\
    &&+(Z_1Z_2Z_3)^{\frac{1}{3}}ds^2_{TN},
\end{eqnarray}
where 
\begin{eqnarray}
&&Z_i=1+\frac{\mu s_i^2 \rho_0}{r_\infty^2 H_i(r_\infty)\rho},\\
&&T=t(H_1(r_\infty)H_2(r_\infty)H_3(r_\infty))^{-\frac{1}{3}},
\end{eqnarray}
and $ds^2_{TN}$ is the metric on self-dual Euclidean Taub-NUT space written in terms of the Gibbons-Hawking coordinate system, which is given by
\begin{eqnarray}
ds^2_{TN}=V^{-1}(d\rho^2+\rho^2(\sigma_1^2+\sigma_2^2))+V\frac{R(r_\infty)}{4}\sigma_3^2
\end{eqnarray}
with
\begin{eqnarray}
V^{-1}=1+\frac{\rho_0}{\rho}.
\end{eqnarray}
This coincides with the extremal three charge black hole solutions in Taub-NUT space~\cite{BW}.

\section*{Acknowledgments} 

ST is supported by the JSPS under Contract No. 20-10616.

\end{document}